\documentstyle[11pt]{article}

\newcommand{\be}{\begin{equation}}
\newcommand{\ee}{\end{equation}}
\newcommand{\lb}{\label}

\begin{document}
\title{Free Decay of Turbulence and Breakdown of Self-Similarity}
\author{Gregory L. Eyink \\{\em Department of Mathematics}\\
{\em University of Arizona}\\{\em Tucson, AZ 85721} \\
and\\
David J. Thomson \\{\em Meteorological Office}\\
{\em London Road, Bracknell, Berkshire}\\
{\em RG12 2SZ, UK}}
\date{ }
\maketitle
\begin{abstract}
It has been generally assumed, since the work of von Karman and Howarth
in 1938, that free decay of  fully-developed turbulence is
self-similar. We present
here a simple phenomenological model of the decay of 3D incompressible
turbulence, which predicts breakdown of self-similarity for
low-wavenumber spectral  exponents $n$ in the range $n_c<n<4$, where
$n_c$ is some threshold wavenumber. Calculations with the
eddy-damped quasi-normal Markovian approximation give  the value
as $n_c\approx 3.45$. The energy spectrum for this range of exponents
develops two  length-scales, separating three distinct wavenumber
ranges.
\end{abstract}

\newpage

The decay of homogeneous, isotropic turbulence is a classical problem.
For a good review, see Lesieur$^1$ (chapter VII, section 10). Since the early
work of von K\'{a}rm\'{a}n and Howarth$^2$, it has often been
assumed that the  decay is self-similar for length-scales outside the
dissipation-range.  Precisely, this assumption means  that the
time-dependent energy spectrum may be written as
\be E(k,t) = v^2(t) \ell(t) F( k\ell(t)), \lb{0} \ee
where $v(t)$ is the rms velocity fluctuation, $\ell(t)$ is the integral
length-scale, and $F(\kappa)$ is a  dimensionless scaling function  (see
section 10.2 of Lesieur$^1$). However, recent studies of two exactly
soluble  models---the Burgers equation$^3$ and the Kraichnan white-noise
passive scalar equation$^4$---have shown that such self-similarity does
not always hold. In particular, Gurbatov {\em{et al.}}$^3$
observed that decaying Burgers turbulence develops {\it two distinct}
length-scales when the low wave number spectral exponent $n$ lies in the
range $1<n<2$. The energy spectrum can then no longer be divided into just a
low-wavenumber range $k\ell(t)\ll 1$ with $E(k,t)\sim A k^n$ and an
inertial range  $k\ell(t)\gg 1$ with $E(k,t)\sim k^{-2}.$  Instead
a new spectral range develops intermediate to
these two with $E(k,t)\sim C(t)k^2,\,\,\,C(t)\propto (t-t_0)^{1/2}
\ln^{-5/4}(t-t_0)$. Since it was the first author of reference 3 who
observed this state of affairs in  Burgers decay$^5$, we call it the
``Gurbatov phenomenon''.

The explanation of this new range  lies in the phenomenon of a $k^2$
backtransfer for Burgers dynamics, the analogue of the
$k^4$ backtransfer  discovered by Proudman and Reid$^6$ in 1954 for the 3D
Navier-Stokes equations. According to traditional beliefs, the
backtransfer term  in Burgers should be overwhelmed at low-wavenumbers
by the original $A k^n$ spectrum, which, for $1<n<2$, is asymptotically
much the larger. This statement, however, ignores the fact that the
coefficient $C(t)$ of the backtransfer term is growing in time,  while
the coefficient $A$ of the lowest wavenumber spectrum $\sim k^n$ is
independent of time. This leads in the Burgers decay to a new
length-scale $\ell_*(t)\gg \ell(t)$, with the $k^2$ backtransfer
spectrum dominating throughout the  intermediate range $k\ell_*(t)\gg 1$
and $k\ell(t)\ll 1$. An analogous ``Gurbatov phenomenon'' was found to
occur  in the Kraichnan passive scalar model$^4$. It is our
purpose to present a similar theory of the breakdown  of
self-similarity for the 3D incompressible Navier-Stokes equations.

Let us consider first the case in which the energy spectrum is dominated by the
$k^4$ backtransfer term
at the lowest wavenumbers (as it will be if the initial spectrum has $n\geq
4$). In this case, one may suppose that
\be E(k,t) \propto \left\{ \begin{array}{ll}
                            C(t)k^4 & k\ell(t)\leq 1 \cr
                            \varepsilon^{2/3}(t) k^{-5/3} & k\ell(t)\geq 1.
                           \end{array} \right. \lb{1} \ee
The continuity of the spectrum at the juncture $k\ell(t)=1$ imposes a relation
\be \varepsilon(t) \propto C^{3/2}(t) \ell^{-17/2}(t). \lb{2} \ee
Two other relations follow by standard Kolmogorov dimensional analysis:
\[ {{dC}\over{dt}}(t)\propto \varepsilon(t)\ell^5(t) \]
and
\[ {{d\ell}\over{dt}}(t) \propto \varepsilon^{1/3}(t)\ell^{1/3}(t). \]
Needless to say, all three of these relations hold in the standard analytical
closures such as the eddy-damped quasi-normal Markovian approximation
(EDQNM). Because we
have three relations and three unknowns ($C,\ell,\varepsilon$), we can find
a solution. In fact, using (\ref{2})
to eliminate $\varepsilon(t)$, we obtain from the other two equations
\be {{dC}\over{dt}}(t)\propto C^{3/2}(t)\ell^{-7/2}(t) \lb{5} \ee
and
\be {{d\ell}\over{dt}}(t) \propto C^{1/2}(t)\ell^{-5/2}(t). \lb{6} \ee
We can then derive from (\ref{5}) and (\ref{6}) that
\[ {{dC}\over{d\ell}} \propto {{C}\over{\ell}} \]
so that, for some arbitrary power $p$,
\be C(t) \propto \ell^p(t). \lb{8} \ee
Unfortunately, it does not seem possible to calculate the precise value of $p$
without making explicit
use of the Navier-Stokes dynamics, either exactly or within a closure
approximation. We shall quote below
the value which follows from EDQNM results. We can only say {\it a priori} that
we expect $p>0$, since
$C(t)$ should grow with time because of the backtransfer. In what follows we
shall not specify $p$,
but work with an arbitrary value.

Eliminating $C(t)$ by substituting (\ref{8}) in (\ref{6}) gives
\[ {{d\ell}\over{dt}}(t) \propto \ell^{{{p-5}\over{2}}}(t), \]
and hence
\be \ell(t) \propto (t-t_0)^{{{2}\over{7-p}}} \lb{10a} \ee
and
\[ C(t) \propto (t-t_0)^{{{2p}\over{7-p}}}. \]
Notice $C(t)\propto \ell^p(t)$ substituted into (\ref{2}) gives also
\[ \varepsilon(t) \propto \ell^{{{3p-17}\over{2}}}(t). \]
The law of the energy decay can then be obtained from $E(t)\propto
(\varepsilon(t)\ell(t))^{2/3}$,
which yields
\be E(t) \propto \ell^{p-5}(t) \propto (t-t_0)^{-{{2(5-p)}\over{7-p}}}. \lb{12}
\ee
Of course, the result $E(t) \propto \ell^{p-5}(t)$ could also be deduced
directly from (\ref{8})
and the spectral model (\ref{1}).  For any choice of $p$, the above spectral
decay law is self-similar.
In fact, using $v^2(t)\ell(t)\propto \ell^{p-4}(t)$, $C(t)\propto \ell^p(t),$
and $\varepsilon^{2/3}(t)\propto \ell^{p-17/3}(t)$,
which follow from the previous relations, we obtain (\ref{0}) with
\[ F(\kappa) \propto \left\{ \begin{array}{ll}
                              \kappa^4 & \kappa\leq 1 \cr
                              \kappa^{-5/3} & \kappa \geq 1.
                           \end{array} \right. \]

The results in (\ref{10a}) and (\ref{12}) may be compared with those
obtained from numerical solution of EDQNM,  $\ell(t)\propto
(t-t_0)^{0.31}$ and $E(t)\propto (t-t_0)^{-1.38}$ (see  section 10.2 of
Lesieur$^1$). From this we may infer  the value $p=0.55$ for EDQNM.
However, the above exponent values do not depend too sensitively on the
precise value of $p$,  which is thus poorly determined by them. For
example, the value $p=1$ implies the relations $\ell(t)\propto
(t-t_0)^{1/3}$  and $E(t)\propto (t-t_0)^{-4/3},$ which are also in
close agreement with the numerical EDQNM results.

The decay laws which follow from this spectral model, for any $p$, are
the same as those which are traditionally believed  to hold in a
self-similar decay for the spectrum with low-wavenumber exponent $n_c=
4-p$ and {\it constant} coefficient $A$.  This constancy of the
coefficient $A$ for $-1<n<4$ is called the {\it permanence of the large
eddies} (PLE)$^7$.  In fact, according to the theory based
upon self-similarity and PLE, it is usually deduced that
\be \ell(t)\propto (t-t_0)^{a},\,\,\,\,\,a={{2}\over{n+3}}, \lb{14} \ee
and
\be E(t)\propto (t-t_0)^{-b},\,\,\,\,\,b={{2(n+1)}\over{n+3}} \lb{15} \ee
for $-1<n<4$ (see section 10.2 of Lesieur$^1$, section
7.7 of Frisch$^7$, or Clark and Zemach$^8$). We find, if we take $n= 4-p$,
that we obtain the same values of
$a$ and $b$ as before. There is of course no contradiction here, because the
coefficient $C(t)$  of the $k^4$ term in (2) depends upon time and
this changes the decay laws.

It is no coincidence that the decay laws for $n=4$ coincide with
those for the standard model based upon  PLE when $n= n_c$. In fact, we
argue, following Gurbatov {\em{et al.}}$^3$ and Eyink and Xin$^4$,
that {\it there is no self-similar
decay at  all when} $n_c<n<4$. We continue to adopt the PLE hypothesis.
However, we question the additional, implicit assumption in deriving
decay laws that $\ell(t)$ is the only length scale in the problem.  We
consider the  consequences of the $C(t)k^4$ backtransfer spectrum, with
$C(t)\propto \ell^p(t)$ as determined above, and consider the
possibility that an intermediate spectral range may form which is
dominated by the backtransfer.  Thus, we take as our model
spectrum
\[ E(k,t) \propto \left\{ \begin{array}{ll}
                            A k^n & k\ell_*(t)\leq 1 \cr
                            C(t)k^4 & k\ell_*(t)\geq 1, k\ell(t)\leq 1 \cr
                            \varepsilon^{2/3}(t) k^{-5/3} & k\ell(t)\geq 1.
                           \end{array} \right. \]
Continuity of the spectrum at the juncture $k\ell_*(t)=1$ requires that
\[ \ell_*^{4-n}(t) \propto C(t)\propto \ell^p(t) \]
and thus
\[ \ell_*(t) \propto \ell^{{{p}\over{4-n}}}(t). \]
The intermediate $k^4$ range only survives---and grows---if $\ell_*(t)\gg
\ell(t)$
for large $t$. Clearly this requires that $p/(4-n) > 1$ or
$n>4-p=n_c$. The growth rate of $\ell_*(t)$ becomes infinitely large as
$n$ approaches $4$ from below, and, in that limit, the $k^4$ region
grows to infinite extent. For $n>4$---as in the  traditional view---the
PLE hypothesis breaks down and the decay is self-similar, governed by
the first model (\ref{1}).

We see that in the range $n_c<n<4$, the decay is not self-similar, as there are
two distinctive length-scales
$\ell(t)$ and $\ell_*(t)$. The $Ak^n$ low-wavenumber spectrum does not dominate
the $C(t)k^4$ backtransfer spectrum
over the whole range $k\ell(t)\ll 1$, because the latter has an increasing
coefficient. Instead, the $k^4$ region is
growing in extent and it is this region which matches onto the energy range at
$k\ell(t)\approx 1$.
It follows that the decay laws are those determined by the backtransfer
spectrum (\ref{1}) {\it over the whole
range of low-wavenumber exponents greater than} $n_c$:
\[ \ell(t)\propto (t-t_0)^{{{2}\over{7-p}}},\,E(t)\propto
(t-t_0)^{-{{2(5-p)}\over{7-p}}}\,\,\,\,\,n>n_c. \]
Thus, the decay laws (\ref{14}), (\ref{15}) hold only for $-1<n<n_c$. In
our opinion, this is a state of affairs far  more {\it a priori}
plausible than the traditionally presented one. In fact, in the
traditional view there is  a monotonic growth in the energy decay
exponent $b= 2(n + 1)/(n + 3)$ over the range $-1<n<4$, which then
suffers  a discontinuous drop at $n=4$. This seems unphysical. In our
theory, the exponent $b$ increases as a function of  $n$ over the range
$-1<n<n_c$, but for larger $n$ sticks at the value for $n=n_c$. Although
the lowest wavenumber  spectrum then does satisfy PLE, it does not match
onto the energy range and it plays no role in the energetics of the
decay. In fact, the energy $E_*(t)$ in the lowest wavenumber range where
PLE holds is
\[ E_*(t)\propto \ell_*^{-(n+1)}(t). \]
Since the total energy scales as $E(t) \propto \ell^{-(n_c+1)}(t)$ for
$n_c=4-p$, there is a negligible fraction   of energy in the PLE range
asymptotically in time for $n>n_c$.  Hence, at these very long times,
self-similarity is  effectively restored, described again by the first
model (\ref{1}).

We presume these facts will also follow from analytical closures, such
as EDQNM, since the basic ingredients  of our phenomenological theory
are already present there. Thus, we expect that those closures have no
self-similar  decay solutions in the range $n_c<n<4$. It is interesting
that the verifications of self-similarity which have been made do not
seem to have included values in this range. See for example figure
VII-8 in Lesieur$^1$, where an impressive similarity collapse is shown,
but only for $n=2$ and $n=4$. We expect that the assumption  of a
self-similar decay in EDQNM-type closures for $n_c<n<4$ will lead to a
realizability violation, similar  to what was found for the Kraichnan
model$^4$. Of course, positive spectra are guaranteed for
EDQNM,  but only if one actually solves the model and not if one makes
hypotheses (such as self-similarity) which may be inconsistent with the
closure equations themselves!

A theory analogous to that presented here can be developed for many other
turbulent decay problems. For example, the case of stationary
turbulence with a Richardson eddy-diffusivity $K(r)\sim r^{4/3}$ was
studied by Eyink and Xin$^4$, within the Kraichnan model.
It was found that a decaying scalar
in the inertial-convective interval experiences self-similar decay
only for low-wavenumber scalar spectral exponents which are initially in the
ranges $-1<n<8/3$ and $n>4$. In this model,  the
time-dependence of the constant $C(t)$ could be evaluated, as
$C(t)\propto (t-t_0)^2$, which allowed the determination  of $n_c= 8/3$.
In the range $8/3<n<4$, the self-similar decay is
nonrealizable and is replaced by a two-scale decay  of the type
described above. This state of affairs presumably holds as well for the
passive scalar advected by an actual  (not synthetic) turbulent
velocity. In fact, one of us$^9$ has constructed a simple model of
the mandoline geometry  often used experimentally to study decay of
temperature fluctuations, and found that there is a low wavenumber $k^4$
spectrum  but with decay exponents the same as for $n=8/3$.
Similar results can be derived also for scalars passively advected  by
turbulence which is itself decaying, and for many other such turbulent
decay problems.

\newpage

\noindent{\bf Acknowledgements:} We wish to thank the Isaac Newton Institute
for its support during the period
this note was written. We also wish to thank many of the participants in the
Turbulence Programme at
the Isaac Newton Institute for their
comments and suggestions, in particular U. Frisch for pointing out an error in
an earlier version.

\section*{References}

\noindent
$^1$M. Lesieur, {\it Turbulence in Fluids}, 3rd ed.
(Kluwer, Dordrecht, 1997).

\noindent
$^2$T. von K\'{a}rm\'{a}n and L. Howarth, ``On the
statistical theory of isotropic turbulence,'' Proc.  Roy. Soc. Lond. A
{\bf 164}, 192-215 (1938).

\noindent
$^3$S. N. Gurbatov, S. I. Simdyankin, E. Aurell, U.
Frisch, and G. T\'{o}th, ``On the decay  of Burgers turbulence,'' J.
Fluid Mech. {\bf 344}, 339-374 (1997).

\noindent
$^4$G. L. Eyink and J. Xin, ``Ideal turbulent decay in the
Kraichnan model of a passive scalar,'' preprint.

\noindent
$^5$U. Frisch, private communication.

\noindent
$^6$I. Proudman and W. H. Reid, ``On the decay of a normally
distributed and homogeneous turbulent velocity  field,'' Phil. Trans.
Roy. Soc. Lond A {\bf 247}, 163-189 (1954).

\noindent
$^7$U. Frisch, {\it Turbulence}. (Cambridge U.P.,
Cambridge, 1995).

\noindent
$^8$T. T. Clark and C. Zemach, ``Symmetries and the approach
to statistical equilibrium in isotropic turbulence,'' Phys. Fluids
{\bf 10}, 2846-2858 (1998).

\noindent
$^9$D. J. Thomson, ``Backwards dispersion of particle pairs
and decay of scalar fields in isotropic turbulence,'' in preparation.

\end{document}